\documentclass[useAMS,referee,usenatbib,usegraphicx]{biom}
\usepackage{hyperref}
\usepackage{empheq}
\usepackage{amssymb}
\usepackage{adjustbox}
\usepackage{relsize}
\usepackage{caption}
\usepackage{booktabs}
\usepackage{multirow}
\usepackage[font=normal,labelfont=bf]{subfig}
\usepackage{xcolor}
\hypersetup{
  colorlinks   = true,
  urlcolor     = blue,
  linkcolor    = blue,
  citecolor    = blue
}

\DeclareMathOperator*{\BX}{\mathbf{X}}

\DeclareMathOperator*{\BYstar}{\mathbf{Y^*}}
\newcommand*{\mux}{\bm{\mu}_{(\BX)\,c}}
\newcommand*{\sigmax}{\Sigma_{(\BX)\, c}}
\newcommand*{\muy}{\bm{\mu}_{(\mathbf{Y^*})\,c}}
\newcommand*{\sigmay}{\Sigma_{(\mathbf{Y^*})\, c}}
\newcommand*{\tran}{\!\mathsf{T}}

\title[Patient stratification in multi-arm trials]{Patient stratification in multi-arm trials: \\a two-stage procedure with Bayesian profile regression}

\author{Yuejia Xu$^{1,*}$\email{yuejia@mrc-bsu.cam.ac.uk}, 
Angela M. Wood$^{2,3}$, and Brian D. M. Tom$^{1}$ \\
$^{1}$MRC Biostatistics Unit, University of Cambridge, Cambridge, U.K \\
$^{2}$Cardiovascular Epidemiology Unit, University of Cambridge, Cambridge, U.K\\
$^{3}$NIHR Blood and Transplant Research Unit in Donor Health and Genomics, Cambridge, U.K}

\begin{document}

%  This will produce the submission and review information that appears
%  right after the reference section.  Of course, it will be unknown when
%  you submit your paper, so you can either leave this out or put in 
%  sample dates (these will have no effect on the fate of your paper in the
%  review process!)

%\date{{\it Received December} 2020. {\it Revised xx} 2021.  {\it Accepted March} 2008.}

%  These options will count the number of pages and provide volume
%  and date information in the upper left hand corner of the top of the 
%  first page as in published papers.  The \pagerange command will only
%  work if you place the command \label{firstpage} near the beginning
%  of the document and \label{lastpage} at the end of the document, as we
%  have done in this template.

%  Again, putting a volume number and date is for your own amusement and
%  has no bearing on what actually happens to your paper!  

%\pagerange{xxxx--xxxx} 
%\volume{xx}
%\pubyear{2021}
%\artmonth{December}

%  The \doi command is where the DOI for your paper would be placed should it
%  be published.  Again, if you make one up and stick it here, it means 
%  nothing!

%\doi{10.1111/j.1541-0420.2005.00454.x}

%  This label and the label ``lastpage'' are used by the \pagerange
%  command above to give the page range for the article.  You may have 
%  to process the document twice to get this to match up with what you 
%  expect.  When using the referee option, this will not count the pages
%  with tables and figures.  

\label{firstpage}

%  put the summary for your paper here

\begin{abstract}
Precision medicine is an emerging field that takes into account individual heterogeneity to inform better clinical practice. In clinical trials, the evaluation of treatment effect heterogeneity is an important component, and recently, many statistical methods have been proposed for stratifying patients into different subgroups based on such heterogeneity. However, the majority of existing methods developed for this purpose focus on the case with a dichotomous treatment and are not directly applicable to multi-arm trials. In this paper, we consider the problem of patient stratification in multi-arm trial settings and propose a two-stage procedure within the Bayesian nonparametric framework. Specifically, we first use Bayesian additive regression trees (BART) to predict potential outcomes (treatment responses) under different treatment options %\textcolor{red}{the responses to different treatments} 
for each patient, and then we leverage Bayesian profile regression to cluster patients into subgroups according to their baseline characteristics and predicted potential outcomes. We further embed a variable selection procedure into our proposed framework to identify the patient characteristics that actively ``drive'' the clustering structure. We conduct simulation studies to examine the performance of our proposed method and demonstrate the method by applying it to a UK-based multi-arm blood donation trial, wherein our method uncovers five clinically meaningful donor subgroups.
\end{abstract}

% 197 words, within the word limits

%  Please place your key words in alphabetical order, separated
%  by semicolons, with the first letter of the first word capitalized,
%  and a period at the end of the list.
%


%  As usual, the \maketitle command creates the title and author/affiliations
%  display 

\maketitle

%  If you are using the referee option, a new page, numbered page 1, will
%  start after the summary and keywords.  The page numbers thus count the
%  number of pages of your manuscript in the preferred submission style.
%  Remember, ``Normally, regular papers exceeding 25 pages and Reader Reaction 
%  papers exceeding 12 pages in (the preferred style) will be returned to 
%  the authors without review. The page limit includes acknowledgements, 
%  references, and appendices, but not tables and figures. The page count does 
%  not include the title page and abstract. A maximum of six (6) tables or 
%  figures combined is often required.''

%  You may now place the substance of your manuscript here.  Please use
%  the \section, \subsection, etc commands as described in the user guide.
%  Please use \label and \ref commands to cross-reference sections, equations,
%  tables, figures, etc.
%
%  Please DO NOT attempt to reformat the style of equation numbering!
%  For that matter, please do not attempt to redefine anything!

\section{Introduction}\label{sec:chap4_intro}

Precision medicine is a paradigm that leverages patient heterogeneity to better inform clinical practice through data-driven, evidence-based, and scientifically-rigorous approaches, and it has gained much popularity recently partly due to the wealth of biological, clinical and epidemiological data now available. In particular, patient stratification based on patients' treatment response profiles and baseline characteristics can greatly facilitate the understanding and characterization of the underlying patient heterogeneity, and thus can be of great interest and importance for a variety of medical applications \citep{sies2019subgroup}.

Our research was motivated by a UK-based, multi-arm trial called INTERVAL \citep{moore2014interval}, which was the first randomized trial to investigate the effect of different inter-donation intervals on blood supply and donor health. In INTERVAL, male donors were randomly assigned to 12-week (standard), 10-week, or 8-week inter-donation intervals, and female donors to 16-week (standard), 14-week, or 12-week inter-donation intervals. The outcomes of this trial include the amount of blood collected, and the number of low hemoglobin deferrals (temporary suspension of donors from giving blood) during the trial period. \textcolor{black}{INTERVAL trial participants were well-characterized at baseline, providing an opportunity to investigate donor stratification and characterize different types of donors. %The availability of the extensive data on outcome and individual donor characteristics from the INTERVAL trial creates an opportunity to investigate donor stratification (based, for example, on donors' covariate profiles and potential responses to different inter-donation intervals), and Such an investigation will facilitate the characterization of different types of donors, 
This can be important for achieving more effective targeted recruitment of donors and increased efficiency of blood collection in the future.} For example, it would be useful to stratify donors into those who have the capacity to donate more often than the current clinical practice (\textcolor{black}{often referred to as ``super donors'' who have high donation capacity and minimal number of deferrals}) and those who tend to be deferred more frequently than the average donors (\textcolor{black}{often referred to as ``brittle donors'' %whose iron stores drain at a greater rate than the average donors after blood donation and 
who struggle more than the average donors to replace iron stores after blood donation}). %``Super donors'' may be encouraged to give blood more frequently when there is a blood shortage or if their blood type is rare or universal, while ``brittle donors'' may be asked to donate less often than the general donor population in order to allow enough time for post-donation iron stores replenishment \citep{nihr2019donorchar}. 

A number of data-driven methods for investigating patient heterogeneity with clinical trial data have been proposed (for a review, see \citet{sies2019subgroup}). However, to our knowledge, most of these methods were developed for two-arm trials, and they have not been extended to the situation with multi-level treatments, %(and thus these methods cannot be directly applied to the INTERVAL data for donor stratification)
\textcolor{black}{despite the popularity of multi-arm trials in clinical practice \citep{baron2013nomultiarm}}.

\textcolor{black}{In this paper, we propose a method for leveraging treatment effect heterogeneity to identify patient subgroups %(according to patients' treatment response profiles and covariate profiles) %with differential treatment responses 
in multi-arm trial settings from a clustering perspective. As a remark, we will use the terms ``subgroup'' and ``cluster'' interchangeably throughout this paper.} 
%We approach the heterogeneity detection problem from a clustering perspective. 
%Our aim is to stratify the heterogeneous population and group individuals into more homogeneous subpopulations/subgroups according to patients' treatment response profiles and covariate profiles. 
In practice, the functional form of the relationship between treatment responses and covariates is typically unknown \textit{a priori}\,. To this end, we leverage Bayesian nonparametric approaches. In particular, our proposed method is a two-stage procedure. In the first stage, we use Bayesian additive regression trees (BART) to predict each patient's potential outcomes (treatment responses) under different treatment options. 
\textcolor{black}{In the second stage, we employ a Bayesian mixture model called ``profile regression'' \citep{molitor2010profile} to cluster patients into subgroups according to their covariate profiles and predicted potential outcomes obtained in the first stage}. %, which links the predicted potential outcomes obtained in the first stage %(i.e. multivariate continuous outcome variable in profile regression) to a set of possibly correlated patient characteristics %(i.e. covariates in profile regression) 
%nonparametrically through cluster memberships. 
\textcolor{black}{We also extend the second stage clustering model to incorporate the variable selection feature, via which covariates that actively ``drive'' the clustering structure can be identified (distinguished from covariates that have similar profiles across all clusters).}
%For the proposed approach, the outcome also informs cluster assignments, and this ensures that the clustering result has clinical utility with regard to our aim. 

%We note that in profile regression, the covariates and the outcome are modeled jointly, and thus both the covariate data and the outcome information influence cluster allocations. In clinical studies, covariate data typically give rise to many different clustering structures, but in the context of precision medicine, interest mainly lies in identifying the structure (and characterizing the patient heterogeneity) that is associated with the outcome on which treatment effects are measured (this outcome is the target outcome of clinical interest). In light of this, the inclusion of the target outcome information in the clustering algorithm can be particularly useful, since it helps mitigate the concerns about identifying clusters that are irrelevant for the specific outcome of clinical interest and will lead to more clinically meaningful clustering results \citep{johnson2020profile,bair2013semiclusterreview}.  

\textcolor{black}{Our method has several desirable properties. First, unlike conventional unsupervised clustering methods, our proposed approach allows the  outcome variable to inform cluster membership, which ensures that resulting clusters are associated with the target outcome of clinical interest and are clinically meaningful. Second, our clustering model is based on the Dirichlet process prior, in which case the number of clusters can be directly inferred from the data, thus sidestepping the difficulty of pre-specifying the number of clusters in classical clustering approaches. Third, the proposed approach can handle correlated patient characteristics (models joint effects and the unit of inference is covariate profile) and correlations do not undermine model performance and interpretability. Finally, our method takes into account %the Bayesian framework allows the quantification of 
the uncertainty associated with the number of clusters and cluster assignments, %across Markov chain Monte Carlo (MCMC) iterations 
and the uncertainty of the ``representative'' clustering can be quantified via model-averaging approaches \citep{molitor2010profile}.}

%the interpretative value of each individual covariate \citep{hennig2013clsuterguide}

%For classical clustering approaches (e.g. K-means, Gaussian mixture models), the number of clusters generally needs to be set \textit{a priori}. Even though multiple criteria have been proposed to determine the number of clusters, there is no consensus on the optimal criterion, and thus specifying the number of clusters can be challenging in practice \citep{tibshirani2001gap}. Our method sidesteps this difficulty by using the Dirichlet process prior, in which case the number of subgroups/clusters can be directly inferred from the data. 

%Another important advantage of our proposed approach is that it can handle correlated covariates. This feature is appealing since variables collected in clinical studies are likely to be correlated, and it would be highly desirable if correlations among covariates do not undermine the model performance and the interpretative value of each individual covariate \citep{hennig2013clsuterguide}. 

%In practice, some covariates available in clinical studies may not play an important role in ``driving'' the clustering structure that is clinically relevant. In this case, it would be useful to identify the covariates that actively ``drive'' the clustering components, and to this end, a variable selection procedure is incorporated into our proposed approach. 

This paper is structured as follows. In Section \ref{sec:chap4_method}, we introduce the statistical framework. %of our proposed method for patient stratification. 
The numerical performance of the method is evaluated by simulation studies in Section \ref{sec:chap4_simu}. An application of the proposed method to the INTERVAL data is presented in Section \ref{sec:chap4_app}. We conclude the paper with a discussion in Section \ref{sec:chap4_diss}.

\section{Methodology}\label{sec:chap4_method}
%We propose a two-stage procedure for patient stratification in multi-arm trial settings. Specifically, in the first stage, the Bayesian additive regression tree (BART) is used to predict each individual's potential outcomes under different treatment options. In the second stage, a Bayesian clustering model called ``profile regression'' is considered for stratifying/clustering patients into subgroups according to their covariate profiles and treatment response profiles.

We consider a clinical trial with $K$ treatments (randomized groups) and $n$ subjects. We let $A\in\mathcal{A}=\{1,\ldots,K\}$ denote the treatment assignment, $\BX\in\mathcal{X}$ denote the covariate data, and $Y\in \mathbb{R}$ denote the observed outcome of interest. The potential outcome under treatment $A=a$ is denoted by $Y^*(a)$. \textcolor{black}{Our proposed two-stage patient stratification method is detailed below in Sections \ref{sec:profile_step_1} and \ref{sec:profile_step_2}. A graphical representation of the proposed approach is presented in Web Appendix A (Web Figure 1).}

\subsection{First stage: BART - predict potential outcomes}\label{sec:profile_step_1}
In the first stage, we fit a flexible regression model for $Y$ given $A$ and $\BX$ in order to get the potential outcome estimates for subjects. We denote the predicted potential outcomes under treatments $1,\ldots,K$ by $\widehat{Y}^*(1),\ldots,\widehat{Y}^*(K)$, respectively. In theory, $\widehat{Y}^*(1),\ldots,\widehat{Y}^*(K)$ can be obtained from the observed data using any supervised machine learning or regression algorithms \citep{kunzel2019metalearner}. We use BART in our implementation \citep{hill2011bayesiannonpara}.

%, since its excellent predictive performance has been reported in various applications

We note that $\widehat{Y}^*(1),\ldots,\widehat{Y}^*(K)$ can also be obtained by fitting a separate model for $Y|\BX$ under each treatment option (``separate-learner'') as opposed to fitting one single model for $Y|A,\BX$ using the data from all subjects (``single-learner'') \citep{kunzel2019metalearner}. These two ways of predicting potential outcomes yield similar results in our numerical studies, and we only present the results based on the ``single-learner'' throughout this paper.

\subsection{Second stage: Bayesian profile regression - clustering}\label{sec:profile_step_2}
\citet{molitor2010profile} proposed a Bayesian mixture model called ``profile regression'', which links the outcome vector to a set of possibly correlated covariates nonparametrically through cluster membership. Profile regression clusters subjects into subgroups by leveraging the Dirichlet process mixture model (DPMM). %As has been discussed in Section \ref{sec:chap4_intro}, 
\textcolor{black}{In profile regression, the covariates and the outcome are modeled jointly, and thus both %the covariate data and the outcome information 
influence cluster allocations. This is appealing, since covariate data typically give rise to many different clustering structures, and the inclusion of the target outcome information in clustering can be particularly helpful for ensuring that the clustering result has clinical utility with regard to our aim \citep{bair2013semiclusterreview}.}  %but in the context of precision medicine, interest mainly lies in identifying the structure (and characterizing the patient heterogeneity) that is associated with the outcome on which treatment effects are measured (target outcome). In light of this, 
%The inclusion of the target outcome information in the clustering algorithm can be particularly useful, since it helps mitigate the concerns about identifying clusters that are irrelevant for the specific outcome of clinical interest and will lead to more clinically meaningful clustering results \citep{johnson2020profile,bair2013semiclusterreview}.  

%the outcome also informs the cluster assignments in profile regression due to the joint modeling of the covariates and the outcome, and this ensures that the clustering result has clinical utility with regard to our aim. 

\textcolor{black}{Our second stage clustering model follows the extension of profile regression to the case with a multivariate normal outcome \citep{johnson2020profile}.} % \citet{liverani2015premium} discussed the applications of profile regression in settings with binary, categorical, or univariate continuous outcomes. \citet{johnson2020profile} extended the profile regression model to the case with a multivariate normal outcome, and we follow their proposed framework when developing our model. 
We use $\BYstar=(\widehat{Y}^*(1),\ldots,\widehat{Y}^*(K))$ to denote %the predicted potential outcome vector, which is 
the multivariate normal response variable in the profile regression model. The likelihood function is given by
\begin{equation}\label{eqn:profile_reg_like}
   p(\BYstar,\BX|\bm{\pi},\bm{\Theta},\bm{\Phi})=\sum_{c=1}^{\infty}\pi_cf(\BYstar|\Theta_c)f(\BX|\Phi_c),
\end{equation}
where $c$ is the index of mixture components, $\bm\pi=(\pi_1,\pi_2,\ldots)$ is a vector of mixture weights, $\bm{\Theta}=(\Theta_1,\Theta_2,\ldots)$ denotes cluster-specific parameters in the density function for $\BYstar$, and $\bm{\Phi}=(\Phi_1,\Phi_2,\ldots)$ denotes cluster-specific parameters in the density function for $\BX$. We note that by construction, DPMM allows infinite components. Therefore, the summation in (\ref{eqn:profile_reg_like}) goes from $1$ to infinity. 

\subsubsection{Specification of model components}\label{sec:pr_model}
In the following, we describe the model for each component of (\ref{eqn:profile_reg_like}), but we will not provide further details on the Bayesian computation aspect in this paper. We refer the interested readers to \citet{liverani2015premium} and \citet{johnson2020profile} for computational details.  

\paragraph{Mixture weights $\pi_c$}
The mixture weights ($\pi_c$) are modeled according to the stick-breaking construction of the Dirichlet process (DP) as follows \citep{sethuraman1994stickbreaking}:
\begin{equation*}
%\begin{split}
%V_c&\sim\text{Beta}(1,\alpha)~~~\alpha>0,\\ \pi_1&=V_1,\\
%\pi_c&=V_c\prod_{r=1}^{c-1}(1-V_r)~~~\text{for }c\geq2,
%\end{split}
\begin{split}
V_c&\sim\text{Beta}(1,\alpha)~~~\alpha>0,\\ \pi_1&=V_1,~~
\pi_c=V_c\prod_{r=1}^{c-1}(1-V_r)~~~\text{for }c\geq2,
\end{split}
 \end{equation*}
 where $V_1,V_2,\ldots$ are independent random variables. 
 Under this representation, $\alpha$ is the concentration parameter of DP, which reflects the dispersion level and controls the number of non-empty clusters implicitly \citep{teh2010dp,hastie2015convergence,fruhwirth2019dp}. We follow \citet{johnson2020profile} and adopt a Gamma prior for $\alpha$.

\paragraph{The model for covariates $\BX$}
The profile regression model can handle both continuous and discrete covariates. In the following, we consider the case where $\BX$ consists of $p_1$ continuous covariates and $p_2$ discrete covariates. In order to describe the complete covariate model, we assume that $p_1,p_2\geq 1$ for now. The situations with $p_1=0$ or $p_2=0$ will be discussed later. We let $\mathbf{X}^{\text{cont}}$ and $\mathbf{X}^{\text{disc}}$ denote the subset of continuous and discrete covariates in $\BX$, respectively. Without loss of generality, and for notational convenience, we assume that the first $p_1$ covariates in $\BX$ are continuous, i.e. $X_j$ (the $j^{\text{\,th}}$ covariate in $\BX$) is continuous for $j={1,\ldots,p_1}$, and $X_j$ is discrete for $j={p_1+1,\ldots,p_1+p_2}$.

We assume that $\mathbf{X}^{\text{cont}}$ and $\mathbf{X}^{\text{disc}}$ are independent conditional on the cluster assignments, and then the ``density'' for covariates $\BX$ can be written as: 
\begin{equation}\label{eqn:profile_reg_x}
  f(\BX|\Phi_c)=f(\mathbf{X}^{\text{cont}}|\mux,\sigmax)f(\mathbf{X}^{\text{disc}}|\bm{\Psi}_c),  
\end{equation}
where $\Phi_c=(\mux,\sigmax,\bm{\Psi}_c)$ represents the cluster-specific parameter set for covariates.
In particular, $\mux$ and $\sigmax$ are the mean vector and the covariance matrix for continuous covariates in cluster $c$, and $\bm{\Psi}_c$ denotes the parameter for discrete covariates in cluster $c$. Note that we add ``$(\BX)$'' in the subscripts of the parameters for continuous covariates (i.e. $\mux$ and $\sigmax$) in order to distinguish them from the parameters for the outcome model. If all the covariates in $\BX$ are continuous (i.e. $p_2$=0), the right-hand side of Equation (\ref{eqn:profile_reg_x}) will be replaced by $f(\BX|\mux,\sigmax)$. On the other hand, if all the covariates in $\BX$ are discrete (i.e. $p_1$=0), the right-hand side of Equation (\ref{eqn:profile_reg_x}) will be replaced by $f(\BX|\bm{\Psi}_c)$.

For $\mathbf{X}^{\text{cont}}$, we assume the following probability density function: 
\begin{small}
\begin{equation}\label{eqn:prof_x_cont}
\begin{split}
  f(\mathbf{X}^{\text{cont}}\big\vert\,\mux,\sigmax)=\Big\{(2\pi)^{\mathlarger{p}_1}\big\vert\sigmax\big\vert\Big\}^{-\frac{1}{2}}\exp{\bigg\{-\frac{1}{2}\big(\mathbf{X}^{\text{cont}}-\mux\big)^{\tran}\Sigma^{-1}_{(\BX)\, c}\big(\mathbf{X}^{\text{cont}}-\mux\big)\bigg\}}. 
  \end{split}
\end{equation}
\end{small}
The conjugate normal-inverse-Wishart (NIW) prior is used for inference, i.e. 
\begin{equation*}
\begin{split}
\mux\,|\,\sigmax&\sim\mathcal {N}(\bm{\mu}_{(\mathbf{X})},\sigmax/\kappa_{(\mathbf{X})}),\\
\sigmax&\sim\mathcal{W}^{-1}(\Lambda_{(\mathbf{X})},\nu_{(\mathbf{X})}),\\
\big(\mux,\sigmax\big)&\sim {NIW}(\bm{\mu}_{(\mathbf{X})},\kappa_{(\mathbf{X})},\Lambda_{(\mathbf{X})},\nu_{(\mathbf{X})}),
 \end{split}
\end{equation*}
 where $\mathcal{N}$ denotes the multivariate normal distribution, $\mathcal{W}^{-1}$ denotes the inverse-Wishart distribution, ${NIW}$ denotes the normal-inverse-Wishart distribution, and $\bm{\mu}_{(\mathbf{X})}$, $\kappa_{(\mathbf{X})}$, $\nu_{(\mathbf{X})}$, and $\Lambda_{(\mathbf{X})}$ (matrix) are hyperparameters. 

For discrete covariates, we assume that they are locally independent (independent conditional on the cluster assignments). Then the probability mass function of $\mathbf{X}^{\text{disc}}$ is given by
\vspace{-1.2em}
\begin{equation}
\label{eqn:prof_x_disc}
f(\mathbf{X}^{\text{disc}}|\bm{\Psi}_c)=\prod_{j={p}_ 1+1}^{p_1+p_2}\psi_{c,j,X_{j~,}}
\end{equation}
where $\psi_{c,j,k}$ denotes the probability that covariate $j$ takes the value $k$ in cluster $c$, $j=p_1+1,\ldots,p_1+p_2$, $k=1,\ldots,K_j$ ($K_j$ denotes the number of categories for covariate $j$). Following \citet{liverani2015premium}, we adopt the conjugate Dirichlet prior, i.e. 
\begin{equation*}
 {\Psi}_{c,j}=(\psi_{c,j,1},\psi_{c,j,2},\ldots,\psi_{c,j,K_j})\sim \text{Dirichlet}(\bm{a}_j),  
\end{equation*}
 where $\bm{a}_j=(a_{j,1},a_{j,2},\ldots,a_{j,K_j})$, $j=p_1+1,\ldots,p_1+p_2$. 

\vspace{2em}
\paragraph{The model for the outcome $\BYstar$}
 We let $\Theta_c=(\muy,\sigmay)$ denote the cluster-specific parameters for the multivariate normal outcome $\BYstar$ and we consider a multivariate normal (MVN) model for $\BYstar$, i.e. 
\begin{small}
\begin{equation*}
 f(\BYstar\big\vert\,\muy,\sigmay)=\Big\{(2\pi)^{K}\big\vert\sigmay\big\vert\Big\}^{-\frac{1}{2}}\exp{\bigg\{-\frac{1}{2}\big(\BYstar-\muy\big)^{\tran}\Sigma^{-1}_{(\BYstar)\, c}\big(\BYstar-\muy\big)\bigg\}}. 
\end{equation*}
\end{small}
\noindent Similar to $\mathbf{X}^{\text{cont}}$, we specify the conjugate normal-inverse-Wishart prior for inference. Specifically,
\begin{equation*}
\begin{split}
\muy\,|\,\sigmay&\sim\mathcal{N}(\bm{\mu}_{(\BYstar)},\sigmay/\kappa_{(\BYstar)}),\\
\sigmay&\sim\mathcal{W}^{-1}(\Lambda_{(\BYstar)},\nu_{(\BYstar)}),\\
\big(\muy,\sigmay\big)&\sim {NIW}(\bm{\mu}_{(\BYstar)}, \kappa_{(\BYstar)},\Lambda_{(\BYstar)},\nu_{(\BYstar)}),
\end{split}
\end{equation*}
where $\bm{\mu}_{(\BYstar)}$, $\kappa_{(\BYstar)}$, $\nu_{(\BYstar)}$ and $\Lambda_{(\BYstar)}$ are hyperparameters. We note that in the profile regression model, we do not impose any specific covariance structure on $\BYstar$, and $\sigmay$ is estimated in a data-driven way.

\subsubsection{Post-processing of the clustering output}\label{sec:post_process}
\textcolor{black}{\paragraph{Identify the ``representative'' clustering}
We note that the proposed Bayesian mixture modeling framework (stochastic) takes into account the uncertainty associated with the number of clusters and cluster assignments, and the clustering output can vary across iterations of the Markov chain Monte Carlo (MCMC) sampler. To facilitate the interpretation of the clustering results, we ``summarize'' the clustering output across iterations and identify a ``representative'' clustering structure.} In each MCMC iteration, we can construct an $n\times n$ score matrix, where the element $(i_1,i_2)$ equals to 1 if individuals $i_1$ and $i_2$ are allocated to the same cluster in this iteration, and equals to 0 if $i_1$ and $i_2$ are assigned to different clusters. We can then average the score matrices over all MCMC iterations to get a posterior similarity matrix, $\mathbf{S}$, which records the probability that two individuals are assigned to the same cluster (i.e. $\mathbf{S}$ records the posterior co-clustering probabilities of all pairs of individuals). The ``representative'' clustering can be identified as the partition of the data that best represents $\mathbf{S}$. Specifically, we use the partitioning around medoids (PAM) algorithm \citep{kaufman1990pam}: PAM is directly applied to the posterior dissimilarity matrix $1-\mathbf{S}$, which allocates individuals to clusters in a way consistent with $\mathbf{S}$. 
For each fixed number of clusters up to a prespecified maximum, we select the best PAM partition that minimizes the sum of dissimilarities between the center of each cluster and all other members of the same cluster. The final ``representative'' clustering is then chosen by maximizing the average silhouette width \citep{rousseeuw1987asd} across these best PAM partitions \citep{molitor2010profile}. 
\textcolor{black}{\paragraph{Quantify the uncertainty associated with the ``representative'' clustering}
We can evaluate how confident we are about the ``representative'' clustering by examining whether or not across different iterations the second-stage clustering model consistently clusters individuals in a way similar to the ``representative'' clustering, and we would expect the credible intervals associated with cluster parameter estimates to be narrower for more consistent clustering (stronger clustering signal). To this end, the model-averaging approach discussed in \citet{molitor2010profile} can be employed under our proposed framework. %For example, for a continuous covariate, at each iteration of the sampler, we can compute the average od 
} 
%We can quantify the uncertainty associated with the ``representative'' clustering of the data and characterize each cluster in the ``representative'' clustering using the model-averaging approach discussed in \citet{molitor2010profile}.} 

\subsubsection{Variable selection in profile regression}\label{sec:vs_profile_reg}
%In many clinical applications, the number of covariates may be quite large and the full covariate profiles can be hard to interpret. It is likely that some of the covariates collected in clinical studies have similar profiles across all clusters and it would be of interest to determine which covariates actively ``drive'' the clustering structure. 
Variable selection methods can be embedded into profile regression in order to identify the covariates that contribute significantly to the formation of clusters. In our implementation, we follow the variable selection approach taken by \citet{liverani2015premium}.

\paragraph{Continuous covariates}
We let $\bm{\gamma}^{\text{cont}}_c=(\gamma_{c,1},\gamma_{c,2},\ldots,\gamma_{c,\,p_1})$, where $\gamma_{c,j}$ is a binary random variable that determines whether or not covariate $j$, $j=1,\ldots,p_1$, is important for allocating subjects to cluster $c$ ($\gamma_{c,j}=1$ if the answer is yes and 0 otherwise). Let $\bar{x}_j$ denote the average value of covariate $j$ (sample average), for $j=1,\ldots,p_1$, and we define $\bm{\mu}^*_c=(\mu^*_{c,1},\mu^*_{c,2},\ldots,\mu^*_{c,\,p_1})$, where 
\begin{equation}
%\begin{split}
%\mu^*_{c,j}&=\gamma_{c,j}\,\mu_{c,j}+(1-\gamma_{c,j})\,\bar{x}_j\\
%&=(\mu_{c,j})^{\gamma_{c,j}}\times {(\bar{x}_j)}^{(1-\gamma_{c,j})}, 
%\end{split}
\begin{split}
\mu^*_{c,j}=\gamma_{c,j}\,\mu_{c,j}+(1-\gamma_{c,j})\,\bar{x}_j
=(\mu_{c,j})^{\gamma_{c,j}}\times {(\bar{x}_j)}^{(1-\gamma_{c,j})}, 
\end{split}
\end{equation}
where $\mu_{c,j}$ is the $j^{\text{\,th}}$ element of $\mux$, $j=1,\ldots,p_1$. We then replace $\mux$ in (\ref{eqn:prof_x_cont}) with $\bm{\mu}^*_c$. 
We assume that $\gamma_{c,j}\sim \text{Bernoulli}(\rho_j)$, $j=1,\ldots,p_1$. A sparsity inducing prior is used for $\rho_j$ \citep{papathomas2012vs,liverani2015premium}.

\paragraph{Discrete covariates}
We can perform variable selection on discrete covariates in a similar manner. 
We let $\bm{\gamma}^{\text{disc}}_c=(\gamma_{c,\,p_1+1},\gamma_{c,\,p_1+2},\ldots,\gamma_{c,\,p_1+p_2})$, where $\gamma_{c,j}$ denotes a binary variable indicating whether or not covariate $j$, $j=p_1+1,\ldots,p_1+p_2$, is important for assigning subjects to cluster $c$. Let $\psi_{c,j,k}$ denote the probability that covariate $j$ takes the value $k$ in cluster $c$, and $\psi_{0,j,k}$ be the observed proportion of covariate $j$ taking the value $k$, $j=p_1+1,\ldots,p_1+p_2$, $k=1,\ldots,K_j$. To incorporate the variable selection feature, the discrete covariate model (\ref{eqn:prof_x_disc}) is modified accordingly to 
\begin{equation*}
\label{eqn:prof_x_disc_vs}
%\begin{split}
%f(\mathbf{X}^{\text{disc}}|\bm{\Psi}_c,\bm{\gamma}^{\text{disc}}_c)&=\prod_{j=p_1+1}^{{p_1+p_2}}\gamma_{c,j}\,\psi_{c,j,X_j}+(1-\gamma_{c,j})\,\psi_{0,j,X_j}\\
%&=\prod_{j=p_1+1}^{{p_1+p_2}}(\psi_{c,j,X_j})^{\gamma_{c,j}}\times (\psi_{0,j,X_j})^{(1-\gamma_{c,j})}.
%\end{split}
\begin{split}
f(\mathbf{X}^{\text{disc}}|\bm{\Psi}_c,\bm{\gamma}^{\text{disc}}_c)=\prod_{j=p_1+1}^{{p_1+p_2}}\gamma_{c,j}\,\psi_{c,j,X_j}+(1-\gamma_{c,j})\,\psi_{0,j,X_j}=\prod_{j=p_1+1}^{{p_1+p_2}}(\psi_{c,j,X_j})^{\gamma_{c,j}}\times (\psi_{0,j,X_j})^{(1-\gamma_{c,j})}.
\end{split}
\end{equation*}

Similar to the continuous covariate case, we assume that $\gamma_{c,j}\sim \text{Bernoulli}(\rho_j)$, and each $\rho_j$ is assigned a sparsity inducing prior, $j=p_1+1,\ldots,p_1+p_2$.

\section{Simulation studies}\label{sec:chap4_simu}
In this section, we evaluate the performance of our proposed method via simulation studies. 
\subsection{Simulation design}
We consider two simulation scenarios. The first scenario corresponds to the case where all covariates are continuous, and we evaluate the performance of the proposed method under different levels of correlations between covariates. \textcolor{black}{In the second scenario, covariates include a mix of continuous and discrete ones and cluster sizes are unequal. We also extend this scenario to evaluate the performance of the variable selection procedure embedded in our method.} We examine two sample sizes: $n=450$ and $n=900$, and we repeat the simulation 100 times for each scenario. Details on the simulation design are given in the following.

\subsubsection{Scenario 1}
\begin{figure}
    \centering
    \captionsetup[subfigure]{oneside,margin={0.8cm,0cm}}
    \subfloat[Scenario 1]{{\includegraphics[width=7.5cm]{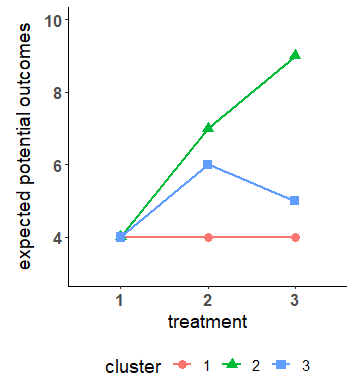} }}%
    \qquad
    \subfloat[Scenario 2]{{\includegraphics[width=7.5cm]{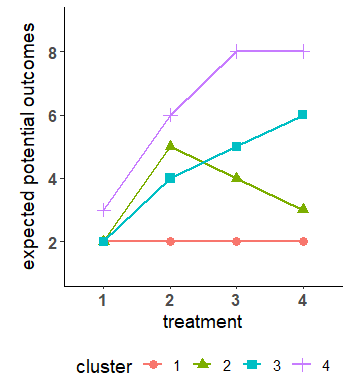} }}%
    \caption{Plot of the expected potential outcome values (by cluster) under different treatments.}%
    \label{fig:sim_setting_outcome}%
\end{figure}

In this setting, we consider a three-arm trial ($K=3$) where treatment $A$ is sampled from \{1,\,2,\,3\} with equal probabilities. $\BX$ consists of 3 normally-distributed continuous covariates: $X_1$, $X_2$, and $X_3$. There are 3 underlying clusters of equal size under this scenario. The mean values of covariates and expected potential outcomes under each treatment option for 3 clusters are summarized in Table \ref{tab:sim_setting} (a). We also plot the expected treatment response profiles of subjects in each cluster in Figure \ref{fig:sim_setting_outcome} (a) for better visualization. %In particular, we can think of the outcome in our simulation studies as a utility score that accounts for the benefit-risk trade-off for each treatment option and without loss of generality, we can assume that the treatment becomes more ``intensive'' as $A$ increases. Cluster 1 corresponds to treatment ``non-responders'', i.e. the potential utility scores remain unchanged regardless of the treatment ``intensity''. Cluster 2 contains subjects whose utility scores get larger as the treatment becomes more ``intensive'', and cluster 3 represents subjects whose optimal treatment is 2 and further increasing the treatment ``intensity'' leads to a lower utility score. 
The observed outcome $Y$ is simulated from a normal distribution with mean $\sum_{a=1}^{3}E\{Y^*(a)\}I(A=a)$ and variance $\sigma^2_Y$. We consider different noise levels in covariates ($\sigma_{\BX}=0.5,1$) and noise levels in the outcome ($\sigma_{Y}=0.5,1$), which reflect different degrees of cluster separability. In addition, we vary the correlation between $X_1$ and $X_2$ ($\rho_{\BX}=0$, 0.5, 0.8 conditional on the cluster assignment; other pairwise correlations are set as 0 conditional on the cluster assignment) to examine the influence of the degree of between-covariate correlations on our proposed method's performance.

\begin{table}
\centering
\subfloat[Scenario 1]{\scalebox{0.8}{%
\begin{tabular}{lccc}
\toprule[0.25ex]
 \rule{0pt}{12pt}
            & cluster 1 & cluster 2 & cluster 3 \\ \hline
       \noalign{\smallskip} 
$E(X_1)$    & 2         & 4         & 6         \\
$E(X_2)$    & 4         & 6         & 1         \\
$E(X_3)$    & 5         & 1         & 3         \\ %\hline
%\noalign{\smallskip} 
\midrule
$E\{Y^*(1)\}$ & 4         & 4         & 4         \\
$E\{Y^*(2)\}$ & 4         & 7         & 6         \\
$E\{Y^*(3)\}$ & 4         & 9         & 5         \\ 
\bottomrule[0.25ex]
\end{tabular}}}
\qquad
\subfloat[Scenario 2]{\scalebox{0.75}{%
\begin{tabular}{lcccc}
\toprule[0.25ex]
 \rule{0pt}{12pt}
            & cluster 1 & \multicolumn{1}{c}{cluster 2} & cluster 3 & cluster 4 \\ \hline
              \noalign{\smallskip} 
$P(X_1=1)$  & 0.2       & \multicolumn{1}{c}{0.4}       & 0.6       & 0.8       \\
$P(X_2=1)$  & 0.4       & 0.4                            & 0.6       & 0.6       \\
$P(X_3=0)$  & 0.1       & 0.2                            & 0.3       & 0.4       \\
$P(X_3=1)$  & 0.15      & 0.3                            & 0.15      & 0.3       \\
$E(X_4)$    & 2         & 4                              & 8         & 6         \\ \midrule
$E\{Y^*(1)\}$ & 2         & 2                              & 2         & 3         \\
$E\{Y^*(2)\}$ & 2         & 5                              & 4         & 6         \\
$E\{Y^*(3)\}$ & 2         & 4                              & 5         & 8         \\
$E\{Y^*(4)\}$ & 2         & 3                              & 6         & 8         \\ 
\bottomrule[0.25ex]
\end{tabular}}}
\caption{Covariate profiles and potential outcome profiles in each cluster.}\label{tab:sim_setting}
\end{table}

\subsubsection{Scenario 2}
In the second setting, we consider a four-arm trial ($K=4$) where treatment $A$ is sampled uniformly from \{1,\,2,\,3,\,4\}. There exist 4 clusters, and the sizes of clusters 1-4 are ${n}/{9}$, ${2n}/{9}$, ${2n}/{9}$, and ${4n}/{9}$, respectively. 
\vspace{-0.5em}
%\paragraph{With no noise covariates}
We consider the situation where there are four covariates, all of which inform the clustering structure. In particular, $X_1$ and $X_2$ are binary variables that take on the value of 0 or 1, $X_3$ is a categorical variable taking on values 0, 1, or 2, and $X_4$ is a normally-distributed continuous variable. In Table \ref{tab:sim_setting} (b), we summarize the covariate profiles and expected treatment response profiles for 4 clusters. The expected treatment response profiles (by cluster) are also plotted in Figure \ref{fig:sim_setting_outcome} (b). %Similar to the first scenario, to facilitate the description of the potential outcome patterns in this setting, the outcome $Y$ can be considered as a utility score that strikes a balance between the benefit and the risk of treatments and we can assume that the treatment becomes more ``intensive'' as $A$ increases: cluster 1 represents treatment ``non-responders''; cluster 2 contains subjects whose ``maximum tolerance'' is treatment 2, and the utility gradually decreases as the treatment becomes more ``intensive'', possibly due to the fact that the increased side-effects outweigh the added benefits by receiving a more ``intensive'' treatment; cluster 3 corresponds to subjects whose response to treatment gets better as the treatment level increases, and cluster 4 represents the ``exceptional-responder'' subgroup given that its members' average responses to all treatment levels are always better than other subgroups. 
The observed outcome $Y$ follows a normal distribution with mean $\sum_{a=1}^{4}E\{Y^*(a)\}I(A=a)$ and variance $\sigma^2_Y$. Four different combinations of noise levels in $X_4$ and noise levels in the outcome $Y$ are considered, namely, $(\sigma_Y,\sigma_{\BX})=(0.2,0.2)$, $(0.5,0.5)$, $(1,0.5)$, $(1,1)$.  \textcolor{black}{We also consider a modification of scenario 2 to evaluate the performance of the proposed variable selection method. %and simulate two noise covariates that do not inform the clustering structure. The performance of the proposed variable selection method (Section \ref{sec:vs_profile_reg}) is evaluated. 
Details are provided in Web Appendix C.}

%\paragraph{With noise covariates}
%To demonstrate the variable selection feature of our proposed method, we modify the simulation setting described in the ``with no noise covariates'' paragraph and include two additional binary covariates, $X_5$ and $X_6$. The profiles of $X_5$ and $X_6$ are the same in all four clusters and thus neither of them contributes to the formation of clusters (i.e. $X_5$ and $X_6$ are noise covariates). Specifically, $P(X_5=1)=0.8$ and $P(X_6=1)=0.15$ for all clusters. The profiles of other covariates and expected treatment responses are the same as those presented in Table \ref{tab:sim_setting} (b). 

%We assess the performance of our proposed method under scenario 2 both when there are no noise covariates and when there are noise covariates. 

\subsection{Evaluation metrics}\label{sec:metrics_describe}
The performance of our proposed method in simulation studies is evaluated based on the following metrics: 
\vspace{-0.5em}
\begin{enumerate}%[label=(\roman*),align=left]
\setlength\itemsep{0.1em}
\item Estimated number of clusters, $\widehat{N}_{\text{cluster}}\,$.
\item Adjusted Rand index (ARI), which measures the similarity between the computed clustering structure and the ground truth (i.e. ideal clustering, we will refer to it as ``class structure'' to distinguish it from the computed clustering structure) \citep{hubert1985ari}. ARI equals to 1 if two partitions of the data are identical and a larger value of ARI indicates a higher level of agreement between two partitions.
\item Conditional entropy-based metrics \citep{rosenberg2007vmeasure}:
\begin{enumerate}
\item Homogeneity, which measures whether all members of a given computed cluster are from the same class (ground truth). Homogeneity score is larger (more desirable) when each cluster contains elements of fewer classes.
\item Completeness, which measures whether all members of a given class (ground truth) are in the same computed cluster. Completeness score is larger (more desirable) when the members of a given class are allocated to fewer clusters.
\end{enumerate}
Both homogeneity and completeness are bounded between 0 and 1.
\end{enumerate}
Details on how to compute ARI, homogeneity, and completeness are provided in Web Appendix B.

\subsection{Simulation results}
In each simulation replicate, BART is run for 6000 iterations to predict the potential outcomes and profile regression is run for 2000 iterations to cluster subjects into subgroups. For both BART and profile regression, the first 1000 iterations are discarded as burn-in. We use the default values specified in the \texttt{BART} package and the \texttt{PReMiuMar} package for all hyperparameters. Investigation of posterior samples suggests no evidence against convergence in our simulation studies.

%(\url{https://github.com/anarouanet/PReMiuMar}) 

\subsubsection{Scenario 1}
Simulation results for scenario 1 are presented in Table \ref{tab:sim_results_scenario_1}.

\begin{table}
\caption[Mean (SD) of adjusted Rand index, completeness, homogeneity, and the estimated number of clusters under scenario 1.]{Simulation results based on 100 replicates: mean (SD) of adjusted Rand index (ARI), completeness, homogeneity, and the estimated number of clusters ($\widehat{N}_{\text{cluster}}$) under scenario 1 for different noise levels in the outcome ($\sigma_{Y}$), noise levels in covariates ($\sigma_{\BX}$), and levels of correlations between covariates $X_1$ and $X_2$ ($\rho_{\BX}$) conditional on the cluster allocation.}
\label{tab:sim_results_scenario_1}
\begin{adjustbox}{width=1.1\textwidth,center}
\begin{tabular}{ccllccccccccc}
\toprule[0.25ex]
                     &                      &        &  & \multicolumn{4}{c}{$n=450$}                                                          &  & \multicolumn{4}{c}{$n=900$}                                                          \\ \cline{5-8} \cline{10-13}   \noalign{\smallskip}
$\sigma_{Y}$               & $\sigma_{\BX}$               & $\rho_{\BX}$ &  & ARI         & Completeness  & Homogeneity & $\widehat{N}_{\text{cluster}}$ &  & ARI         & Completeness & Homogeneity & $\widehat{N}_{\text{cluster}}$ \\ \hline 
  \noalign{\smallskip}
\multirow{3}{*}{0.5} & \multirow{3}{*}{0.5} & 0      &  & 0.99 (0.01) & 0.96 (0.03) & 1.00 (0.00)  & 4.11 (0.96)   &  & 0.97 (0.02) & 0.92 (0.04) & 1.00 (0.00)  & 5.11 (0.87)   \\
                     &                      & 0.5    &  & 0.98 (0.02) & 0.95 (0.04) & 1.00 (0.01)  & 4.17 (1.02)   &  & 0.97 (0.02) & 0.93 (0.04) & 1.00 (0.00)  & 5.00 (0.96)   \\
                     &                      & 0.8    &  & 0.98 (0.02) & 0.95 (0.04) & 1.00 (0.01)  & 4.17 (1.01)   &  & 0.97 (0.02) & 0.93 (0.04) & 1.00 (0.00)  & 5.07 (0.98)   \\  \midrule
\multirow{3}{*}{1}   & \multirow{3}{*}{1}   & 0      &  & 0.88 (0.07) & 0.81 (0.08) & 0.95 (0.02)  & 4.43 (0.88)   &  & 0.77 (0.09) & 0.69 (0.07) & 0.95 (0.01)  & 5.86 (0.94)   \\
                     &                      & 0.5    &  & 0.88 (0.08) & 0.82 (0.09) & 0.97 (0.02)  & 4.44 (1.03)   &  & 0.80 (0.09) & 0.73 (0.08) & 0.96 (0.01)  & 5.76 (1.00)   \\
                     &                      & 0.8    &  & 0.90 (0.08) & 0.84 (0.09) & 0.97 (0.02)  & 4.26 (0.77)   &  & 0.83 (0.08) & 0.75 (0.07) & 0.97 (0.01)  & 5.60 (0.95)   \\ \bottomrule[0.25ex]
\end{tabular}
\end{adjustbox}
\end{table}

The results are fairly robust across different levels of correlations ($\rho_{\BX}$) among covariates. This is a highly desirable property of our proposed approach compared to many other standard approaches, whose performance can be very sensitive to the degree of correlations among covariates due to the multicollinearity problem \citep{molitor2010profile}. 

The $\widehat{N}_{\text{cluster}}$ column shows that our proposed method over-estimates the true number of clusters (${N}_{\text{cluster}}=3$ in this scenario), especially when the sample size is large ($n=900$). This is not surprising given that the profile regression adopts a Dirichlet process mixture model (DPMM): it has been demonstrated that when the true number of clusters is finite and small, the posterior inference on the number of clusters by using the DPMM may be inconsistent, and DPMM tends to over-estimate the true number of clusters and produce some small extraneous clusters around the true components \citep{onogi2011DPncluster,miller2013dpover,yang2019dpncluster,lu2018dp}. This phenomenon is referred to as ``over-clustering'' by \citet{lu2018dp}, which might be due to the sensitivity of DPMM to even minor deviations that exist among clusters. As has been discussed in Section \ref{sec:pr_model}, the choice of the concentration parameter $\alpha$ implicitly affects the expected number of (non-empty) clusters. Results presented in Table \ref{tab:sim_results_scenario_1} are obtained by assuming a Gamma(2,1) prior for $\alpha$ (i.e. the default in the \texttt{PReMiuMar} package; $E(\alpha)=2$). We also examine the empirical results obtained under other commonly-used priors for $\alpha$, for example, Gamma(2,4) ($E(\alpha)=0.5$) suggested by \citet{escobar1995concentration}, and the prior that is matched to the sparse finite mixture model \citep{fruhwirth2019dp}. The estimated number of clusters, $\widehat{N}_{\text{cluster}}$, are very similar across different prior specifications for $\alpha$ (not presented in the paper), even though based on theoretical considerations, the results might differ \citep{fruhwirth2019dp}.

%We note that in order to demonstrate the ``over-clustering'' problem, we do not set an upper bound on the number of clusters or place a constraint on the minimum cluster size in our simulation studies. However, in practice, if the examination of the output suggests that some resulting clusters contain only a small number of subjects, or if we have \textit{a priori} knowledge on the maximum number of subgroups that we are aiming for, such information can be incorporated into the post-processing step straightforwardly in the implementation (e.g. by specifying a value for ``\texttt{maxNClusters}'' in the ``\texttt{calcOptimalClustering}'' function).

Not surprisingly, homogeneity is higher (better) than completeness in most cases, 
given that $N_{\text{cluster}}$ is typically over-estimated by our method. In this sense, it is more likely that subjects who are clustered together are from the same class (the ground truth in simulation studies), and thus leading to ``more homogeneous'' clusters. On the other hand, small extraneous clusters (centered around true clusters) that are produced by DPMM may explain the ``less complete'' clustering results. 

As expected, higher noise levels in covariates ($\sigma_{\BX}$) and outcomes ($\sigma_Y$) result in worse clustering performance (i.e. lower ARI, completeness, and homogeneity, and a larger upward bias in the estimation of ${N}_{\text{cluster}}$), given that we would expect the underlying clustering structure to be less clear (lower cluster separability) with larger values of $\sigma_{\BX}$ and $\sigma_Y$. 

A comparison of the results obtained when $n=450$ with those obtained when $n=900$ suggests that larger sample sizes do not seem to improve the performance of our method in this setting. When the noises in covariates and the outcome are large ($\sigma_Y=\sigma_{\BX}=1$), ARI and completeness even get worse as the sample size increases. One possible reason for this observation is that the over-estimation of $N_{\text{cluster}}$ by DPMM is more pronounced when $n=900$, and this has a subsequent (negative) effect on ARI and completeness.

\begin{figure}
    \centering
    \captionsetup[subfigure]{oneside,margin={0.8cm,0cm}}
    \subfloat[Covariates]{{\includegraphics[width=0.65\textwidth]{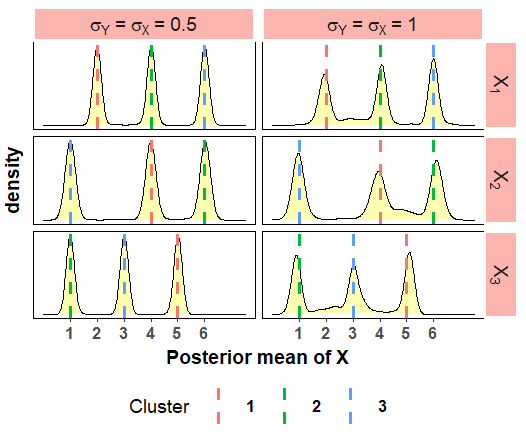} }}%
    \qquad
    \subfloat[Potential outcomes under different treatments]{{\includegraphics[width=0.65\textwidth]{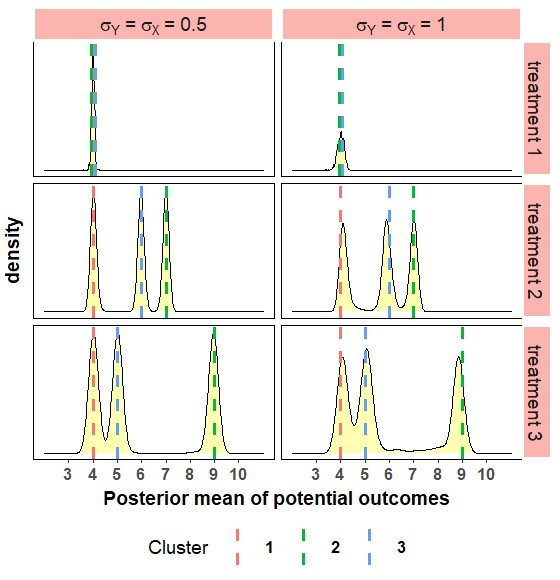} }}%
    \caption{Density plots of cluster-specific mean parameters in scenario 1 with $n=900$ and $\rho_{\BX}=0.5$. Dashed lines refer to true values in each cluster. The left and right panels correspond to the low and high noise settings, respectively.}%
    \label{fig:scen1_density}%
\end{figure}

In addition to measuring the overall clustering accuracy of our proposed method based on the clustering performance metrics described in Section \ref{sec:metrics_describe}, we also examine how well our method performs in terms of recovering the true underlying cluster-specific parameters (e.g. means). To this end, in each simulation replicate, we first obtain the posterior summaries of mean parameters for $X_1$, $X_2$, $X_3$, $Y^*(1)$, $Y^*(2)$, and $Y^*(3)$ in each resulting cluster, and these cluster-specific parameters are then re-weighted by the cluster sizes. We note that the re-weighting step is important: posterior summary statistics for larger clusters should be assigned more weight since larger clusters carry more information. The densities of the re-weighted results over 100 simulation replicates in low ($\sigma_Y=\sigma_{\BX}=0.5$) and high ($\sigma_Y=\sigma_{\BX}=1$) noise settings are plotted in Figure \ref{fig:scen1_density} ((a) for covariates $X_1$, $X_2$, and $X_3$ and (b) for potential outcomes under each treatment option). These plots correspond to the case with $n=900$ and $\rho_{\BX}=0.5$. 
Density plots corresponding to other $n$ and $\rho_{\BX}$ values look similar (do not alter our conclusions) and are thus omitted.

%\begin{figure}
%\centering
%\includegraphics[width=0.95\textwidth]{density_covariates.png}
%\caption[Density plots of cluster-specific mean parameters for profile variables in scenario 1.]{Density plots of cluster-specific mean parameters for $\BX$ in scenario 1 with $n=900$ and $\rho_{\BX}=0.5$. Dashed lines refer to true values of $E(X_1)$, $E(X_2)$, and $E(X_3)$ in each cluster. The left and right panels correspond to the low and high noise settings, respectively.}\label{fig:scen1_density_covariates}
%\end{figure}
%
%\begin{figure}
%\centering
%\includegraphics[width=0.95\textwidth]{density_outcomes.png}
%\caption[Density plots of cluster-specific mean parameters for potential outcomes in scenario 1.]{Density plots of cluster-specific mean parameters for potential outcomes under treatments 1, 2, and 3 in scenario 1 with $n=900$ and $\rho_{\BX}=0.5$. Dashed lines refer to true values of $E\{Y^*(1)\}$, $E\{Y^*(2)\}$, and $E\{Y^*(3)\}$ in each cluster. The left and right panels correspond to the low and high noise settings, respectively.}\label{fig:scen1_density_outcomes}
%\end{figure}

Figure \ref{fig:scen1_density} implies that the true cluster-specific means (dashed lines) for all covariates and potential outcomes can be recovered by our proposed method, even though the estimated number of clusters is greater than the truth. 
When the noise level is high ($\sigma_Y=\sigma_{\BX}=1$), the densities are flatter (nosier) compared to the case with low noise levels ($\sigma_Y=\sigma_{\BX}=0.5$), but our method still manages to recover the truth.

\subsubsection{Scenario 2}
%\paragraph{With no noise covariates}
Simulation results under scenario 2 are summarized in Table \ref{tab:sim_results_scenario_2}. 

\begin{table}
\caption[Mean (SD) of adjusted Rand index, completeness, homogeneity, and the estimated number of clusters under scenario 2 with no noise covariates.]{Simulation results based on 100 replicates: mean (SD) of adjusted Rand index (ARI), completeness, homogeneity, and the estimated number of clusters ($\widehat{N}_{\text{cluster}}$) under scenario 2 with no noise covariates for different noise levels in the outcome ($\sigma_{Y}$) and noise levels in the continuous covariate $X_4$ ($\sigma_{\BX}$).}
\label{tab:sim_results_scenario_2}
\begin{adjustbox}{width=1.1\textwidth,center}
\begin{tabular}{cclccccccccc}
\toprule[0.25ex]
       &        &  & \multicolumn{4}{c}{$n=450$}                                                                                                                                                                                                                                    &  & \multicolumn{4}{c}{$n=900$}                                                                                                                                                                                                                                    \\ \cline{4-7} \cline{9-12} \noalign{\smallskip}
$\sigma_{Y}$ & $\sigma_{\BX}$ &  & ARI         & Completeness & Homogeneity & $\widehat{N}_{\text{cluster}}$ &  & ARI         & Completeness & Homogeneity & $\widehat{N}_{\text{cluster}}$ \\ \hline\noalign{\smallskip}
0.2    & 0.2    &  & 0.99 (0.02) & 0.99 (0.02) & 0.99 (0.03)  & 4.67 (0.83)   &  & 0.99 (0.01) & 0.98 (0.02) & 1.00 (0.00)  & 5.59 (0.91)   \\
0.5    & 0.5    &  & 0.84 (0.06) & 0.79 (0.05) & 0.89 (0.06)  & 5.97 (1.11)   &  & 0.84 (0.05) & 0.75 (0.04) & 0.91 (0.02)  & 7.28 (1.20)   \\
1      & 0.5    &  & 0.81 (0.09) & 0.78 (0.07) & 0.86 (0.06)  & 5.69 (1.24)   &  & 0.82 (0.07) & 0.75 (0.06) & 0.91 (0.02)  & 6.92 (1.13)   \\
1      & 1      &  & 0.40 (0.07) & 0.45 (0.06) & 0.50 (0.06)  & 5.04 (1.50)   &  & 0.37 (0.07) & 0.42 (0.05) & 0.53 (0.03)  & 6.08 (1.34)   \\ \bottomrule[0.25ex]
\end{tabular}
\end{adjustbox}
\end{table}

As in the first scenario, the homogeneity score is almost always higher than the completeness score due to ``over-clustering''. When the noise level is low ($\sigma_Y=\sigma_{\BX}=0.2$), the clusters are well-separated and our proposed method performs almost perfectly in terms of ARI, completeness and homogeneity, despite the over-estimation of $N_{\text{cluster}}$ remaining a problem (for this scenario, $N_{\text{cluster}}=4$). As the noise level increases, the clustering performance gets worse, and it seems to be more sensitive to the change in $\sigma_{\BX}$ than to the change in $\sigma_{Y}$ (when we compare the results corresponding to $(\sigma_Y,\sigma_{\BX})=(1,0.5)$ with those corresponding to $(\sigma_Y,\sigma_{\BX})=(1,1)$). We also observe that $\widehat{N}_{\text{cluster}}$ first increases and then decreases as the noise gets larger, possibly because our proposed algorithm merges some clusters when the true underlying clustering structure has very low separability (i.e. very large noise).

\section{Application to the INTERVAL trial}\label{sec:chap4_app}
We apply our proposed patient stratification approach to the data from the INTERVAL trial. The purpose of this analysis is to uncover donor subgroups with different baseline characteristics and potential ``treatment'' (inter-donation interval in the blood donation context) response profiles. Specifically, we focus our analysis on a ``much-in-demand but vulnerable'' donor population of 884 female donors who were younger than 40 (more at risk of iron deficiency after donating blood) and had O negative blood type (the ``universal'' blood group that can be transfused to any patient in need and used in medical emergencies). The three randomized groups for female donors are 16-week, 14-week, and 12-week inter-donation intervals. The target outcome of our interest is a utility score (denoted by $U$) that accounts for the trade-off between the total units of blood collected per donor over the 2-year trial period (the benefit outcome, denoted by $G$) and the number of low hemoglobin (Hb) deferrals per donor during the same period (the risk outcome, denoted by $R$), i.e. $U=G-b\times R$, where $b$ is the trade-off parameter reflecting the equivalent benefit loss for one unit increase in the risk. We examine the case with $b=3$ to reflect the potential costs of reduced efficiency of blood collection and reduced donor retention due to low Hb deferrals. Seven baseline donor characteristics are considered, including the Short Form Health Survey version 2 (SF-36v2) physical component score (PCS), mental component score (MCS), ferritin level, red blood cell count (RBC), mean corpuscular volume (MCV), mean corpuscular hemoglobin (MCH), and body mass index (BMI). 

In our analysis, we use the default setups that are specified in the \texttt{BART} package and the \texttt{PReMiuMar} package for priors and hyperparameters. We run BART for 6000 MCMC iterations with an initial burn-in of 1000 iterations to predict potential outcomes, and run profile regression for 40000 iterations with a burn-in of 10000 iterations in the clustering step. We do not find strong evidence against convergence. \textcolor{black}{More details on convergence diagnostics are given in Web Appendix D}.  

Unlike in the simulation studies where we can evaluate the performance of our method by calculating external metrics such as ARI, homogeneity, and completeness, in the real data application, the ground truth is not known and thus these external validation metrics cannot be used. Instead, we focus on the interpretation of the ``representative'' clustering of 884 female donors in order to assess whether or not our proposed approach can give insights into donor heterogeneity and stratify donors into clinically meaningful subgroups.

Inspection of the raw output from the profile regression model (before applying the post-processing method described in Section \ref{sec:post_process}) suggests that at each MCMC iteration, there are either 6 or 7 resulting non-empty clusters, with 5 moderately-sized clusters and 1 or 2 very small clusters. In particular, for all MCMC iterations, one of the resulting clusters contains only 1 donor whose ``red blood cell-related'' blood measurements are fairly extreme (lowest RBC, highest MCH and $3^{\text{rd}}$ highest MCV among all 884 female donors). %, and she may suffer from Vitamin B12 or folate deficiency anemia. 
The other small cluster (if it exists, i.e. when the total number of non-empty clusters is 7) includes donors with extremely low PCS (the size of this cluster varies slightly across different iterations, but it is always less than 8).

We apply the post-processing method discussed in Section \ref{sec:post_process} to identify the ``representative'' clustering based on posterior samples. The heatmap of the posterior similarity matrix for 884 female donors is presented in Web Appendix E (Web Figure 4). The ``representative'' clustering consists of 5 donor subgroups, and the sizes of these subgroups are 171, 126, 101, 93, and 393, respectively. This indicates that the 1 or 2 very small cluster(s) in the raw output are merged into larger clusters in this post-processing step.% (the interpretation of the ``merge'' will follow shortly).

%\begin{figure}[h]
%\centering
%\includegraphics[width=0.68\textwidth,height=9cm]{heatmap_initial_30_rep_1_new}
%\caption[Heatmap of 884 female donors' posterior similarity matrix based on profile regression when the utility score is the target outcome.]{Heatmap of 884 female donors' posterior similarity matrix (across 30000 posterior draws) based on the profile regression model when the utility score with $b=3$ is the target outcome. Different colours correspond to different degrees of individuals' similarity to each other (yellow - low similarity; red - high similarity).}\label{fig:interval_heatmap_utility}
%\end{figure}

\begin{figure}
\subfloat[Covariate profiles from the profile regression model: posterior distributions of the mean parameters for clusters 1-5. The red-colored and the blue-colored boxes indicate that the 90\% credible intervals for the cluster-specific mean are above and below the average values across clusters 1-5, respectively. The green-colored boxes imply that the 90\% credible intervals for the cluster-specific mean include the average. We note that the boxes cover interquartile range (lower hinge: the 25$^{\text{th}}$ percentile; upper hinge: the 75$^{\text{th}}$ percentile).]{%
  \includegraphics[clip,width=1\columnwidth,height=7.5cm]{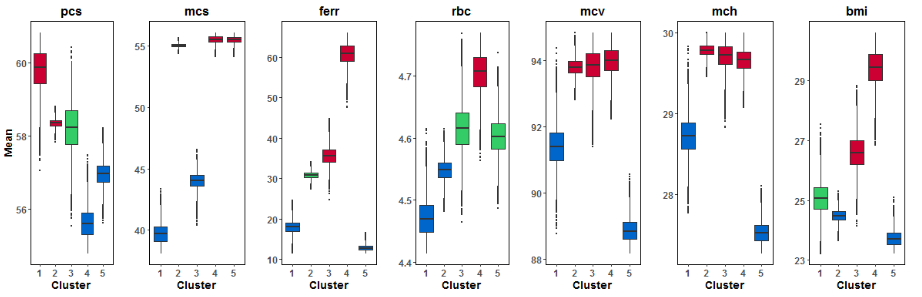}%
}\\[4ex]

\subfloat[Potential outcome profiles from the profile regression model: posterior distributions of the mean parameters for clusters 1-5 under the 16-week (pink), 14-week (green) and 12-week (blue) inter-donation intervals.]{%
  \includegraphics[clip,width=1\columnwidth,height=7.5cm]{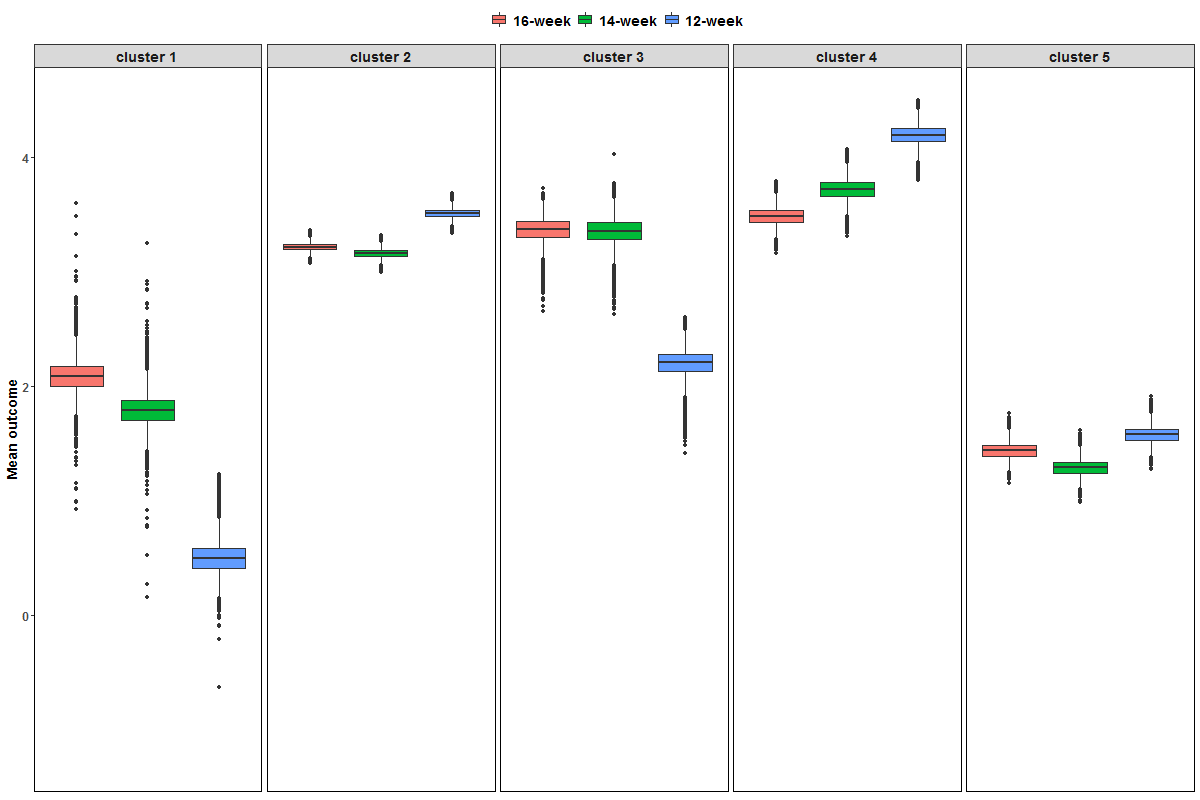}%
}

\caption[Profile regression plots obtained by applying our proposed patient stratification method to the data from 884 female donors in the INTERVAL trial: posterior distributions of the parameters for the ``representative'' clustering.]{Profile regression plots obtained by applying our proposed patient stratification method to the data from 884 female donors who were younger than 40 and had O negative blood type in the INTERVAL trial: posterior distributions of the parameters associated with the response (utility score) and covariates (baseline characteristics) for the ``representative'' clustering.}\label{fig:prof_reg_interval_utility}

\end{figure}

Figure \ref{fig:prof_reg_interval_utility} shows the covariate profiles (posterior distributions of the mean parameters for donors' baseline characteristics) and the potential outcome profiles (posterior distributions of the mean parameters for potential outcomes under 16-, 14-, and 12-week inter-donation intervals) corresponding to each of the five donor subgroups. These results reveal considerable evidence for the presence of heterogeneity within the donor population under investigation. 

Cluster 4 (with 126 donors) represents the potentially ``super donor'' subgroup who  can give blood more frequently than the current clinical practice and for these donors, more frequent donations lead to larger utilities (Figure \ref{fig:prof_reg_interval_utility} (b)). Comparing across subgroups, this group of donors has higher donation capacity since the utility scores are on average higher, especially under the 14-week and 12-week inter-donation intervals. These ``super donors'' are characterized by having high levels of ferritin, RBC, BMI, MCS, MCV and MCH and low levels of PCS. In particular, the most distinguishing feature that differentiates this subgroup from the remaining subgroups is the ferritin level (significantly higher in this subgroup). This is consistent with \textit{a priori} expectation that donors with higher ferritin levels are in general more capable of donating blood more often (especially for female donors).

Cluster 2 is the largest subgroup with 393 donors. Figure \ref{fig:prof_reg_interval_utility} (b) suggests that donors in this subgroup have the potential to donate blood every 12 weeks, even though the gain in the utility score by switching from the 16-week inter-donation interval to the 12-week inter-donation interval is only moderate and not as significant compared to cluster 4. %If there is a shortage of blood, donors in cluster 4 may be encouraged to donate more frequently first, followed by donors in cluster 2. 
Consistent with cluster 4, donors in cluster 2 on average have high MCS, MCV, and MCH. However, their ferritin levels, RBC and BMI are lower than those of donors in cluster 4 (Figure \ref{fig:prof_reg_interval_utility} (a)). 

Cluster 1 (with 101 donors) represents a potentially ``brittle donor'' subgroup. %Donors in this subgroup cannot donate more frequently than every 16 weeks and the 
Utility score gets smaller as the inter-donation interval gets shorter (Figure \ref{fig:prof_reg_interval_utility} (b)). The low ferritin levels, RBC and MCS of these donors (Figure \ref{fig:prof_reg_interval_utility} (a)) may potentially explain why they are vulnerable. As an aside, the cluster with only one donor in the raw output before post-processing is merged into this ``brittle donor'' subgroup. This is sensible since blood-based measurements of this donor suggest that she may be ``vulnerable''.%suffer from anemia and is thus ``vulnerable''.  

Cluster 5 (with 171 donors) also represents a subgroup of donors with low donation capacity. The utility scores of donors in this subgroup are similar under three inter-donation interval options and they are lower than the utility scores of donors in clusters 2, 3, and 4 (Figure \ref{fig:prof_reg_interval_utility} (b)). Donors in cluster 5 (on average) have a high MCS similar to that of donors in cluster 4. However, low values of donation capacity-related characteristics such as ferritin levels, MCV, MCH, and BMI (Figure \ref{fig:prof_reg_interval_utility} (a)) may lead to higher than average deferral rates and lower than average utility scores in this subgroup.  

The utility scores of donors in cluster 3 (with 93 donors) are almost identical under 16-week and 14-week inter-donation intervals but decrease significantly when the donation frequency is every 12 weeks (Figure \ref{fig:prof_reg_interval_utility} (b)). %Therefore, we would assign these donors to the 16-week inter-donation interval, given that if the utility scores under two inter-donation interval options are similar, the less ``intensive'' inter-donation interval would be more desirable in reducing the risk of low Hb deferrals. 
The ferritin levels, MCV, MCH, and BMI are relatively high (Figure \ref{fig:prof_reg_interval_utility} (a)) for these donors, and the drop in the utility score when the inter-donation interval is 12-week may be attributed to the low MCS. %(it is likely that frequent donations demotivate donors with lower MCS from returning to future donation sessions to a higher degree). 

To summarize, our analysis of the INTERVAL data using the proposed approach reveals clinically meaningful donor subgroups within the ``much-in-demand but vulnerable'' donor population under investigation. %These results also demonstrate the potential gain of investigating ``patient'' stratification in addition to estimating the optimal individualized treatment rule (ITR), which is another important research area in precision medicine \citep{zhou2017rwl}. Of course, based on the estimated optimal ITR, $\widehat{\mathcal{D}^*}(\Bx)$, we can stratify the donor population into three subgroups where subgroups 1, 2, and 3 represent donors with $\widehat{\mathcal{D}^*}(\Bx)=$ 16-week, $\widehat{\mathcal{D}^*}(\Bx)=$ 14-week, and $\widehat{\mathcal{D}^*}(\Bx)=$ 12-week, respectively. However, such a stratification does not incorporate all the covariate information (even if covariates are included in the ITR estimation model) and the complete patterns of potential outcomes under all inter-donation interval options, and thus the granularity will be low. For example, the stratification may fail to capture the heterogeneity in the baseline characteristics of donors whose estimated optimal inter-donation intervals are the same. In addition, even though all donors in subgroup 3 achieve the highest utility scores when assigned to the 12-week inter-donation interval, their complete potential outcome patterns might differ (e.g. outcome patterns of some donors may be consistent with those of cluster 2 in Figure \ref{fig:prof_reg_interval_utility} (b) and outcome patterns of other donors may be consistent with those of cluster 4 in Figure \ref{fig:prof_reg_interval_utility} (b)). In these situations, a further split of the three subgroups obtained solely based on $\widehat{\mathcal{D}^*}(\Bx)$ may offer additional insights into the underlying donor heterogeneity. On the other hand, our proposed method has a higher ``resolution'' and enables us to get a more refined stratification of the donor population directly by the joint modeling of the ``imputed'' multivariate outcome (a vector of the potential outcome under each inter-donation interval) and the covariates. We note that if some of the resulting subgroups are fairly similar from the clinical perspective (in terms of clinical meaningfulness), they can be merged \textit{a posteriori}. From our point of view, this would be preferable to starting with a method of low ``resolution'', in which case some useful information on the population heterogeneity may be overlooked. 

\section{Discussion}\label{sec:chap4_diss}

The uncovering of subgroups from a heterogeneous population plays an important role in precision medicine applications. In this paper, we present a two-stage patient stratification approach that leverages Bayesian nonparametric techniques. Our proposed method captures the heterogeneity in the underlying population and clusters the population into subgroups of subjects who share similar covariate profiles and display similar treatment responses. Specifically, in the first stage, we predict the potential outcome under each treatment arm for each patient, and in the second stage, we apply profile regression to link the multivariate potential outcome vector to a set of covariates (can be continuous, discrete, or a mix of continuous and discrete ones) through cluster membership \citep{molitor2010profile}. Based on the posterior samples, the resulting subgroups can be characterized in terms of covariate and treatment response profiles.    

Our method offers several advantages. Firstly, while most existing methods for subgroup identification only cover two-treatment cases, our proposed approach is applicable to multi-arm trials. Secondly, the use of the Dirichlet process prior allows the number of clusters to be estimated from the data, thus bypassing the need for pre-specifying it. Thirdly, our method can properly handle correlated covariates (avoiding well-known problems caused by multicollinearity), which are common in clinical studies. Fourthly, a variable selection procedure is embedded into our model for identifying important covariates that actively ``drive'' the clustering structure (i.e. contribute significantly to the cluster patterns). Lastly, the proposed approach is built under the Bayesian framework and takes into account model uncertainties \citep{molitor2010profile}.%allows the direct quantification of the uncertainties associated with the final clustering results (i.e. the optimal partition) and the corresponding group parameters \citep{molitor2010profile}. 

The application of our proposed method to a subset of the INTERVAL data (a ``much-in-demand but vulnerable'' donor population) identifies 5 clinically meaningful donor subgroups with different donation capacities and covariate (donors' baseline characteristics) profiles. These results provide insight into the underlying donor heterogeneity by highlighting the differences between donors in terms of both baseline characteristics and potential response (to three different inter-donation intervals) patterns, and can be leveraged to inform and guide targeted donor recruitment and donor management strategy. For example, donors who are identified as belonging to the ``super donor'' subgroup may be asked to give blood more frequently if there is a blood shortage or if their blood group is rare or universal. In contrast, donors who belong to the ``brittle donor'' subgroup will be allowed longer time between donations to ensure donor health and safety \citep{nihr2019donorchar}. 

%As has been discussed in Section \ref{sec:profile_step_1}, although we use BART in the first stage of our method to ``impute'' the potential outcome under each treatment option, the validity of our approach does not rely on any particular choice of the outcome imputation model, and BART can be replaced by other sufficiently flexible models (e.g. super learner proposed by \citet{vanderlaan2007superlearner}).

The use of the Dirichlet process prior in the second-stage profile regression model allows the number of clusters to be discovered in a data-driven way. However, as demonstrated by our simulation studies, Dirichlet process mixture model (DPMM) tends to over-estimate the number of clusters (produce some superfluous small-sized clusters). Indeed, the inconsistent inference (over-estimation) on the number of components by DPMM is a well-known problem (referred to as ``over-clustering'' in \citet{lu2018dp}) that has been empirically observed and reported in the literature before, and it appears that in most cases, the extra clusters are centered around the true components and only include a very small number of individuals \citep{onogi2011DPncluster,miller2013dpover}. To our knowledge, how to correct for such inconsistency remains an open question in the field of Bayesian mixture modeling \citep{miller2013dpover,yang2019dpncluster}. %\citet{guha2019nclust} recently proposed a truncation method to resolve the issue, but the success of their approach requires the provision of certain information (e.g. the lower bound for the ratios of the clusters) that is typically unknown in practice \citep{yang2019dpncluster}. %Another proposal is to reduce ``over-clustering'' by penalizing small clusters \citep{lu2018dp}. However, as can be expected, this approach requires careful tuning of the parameter that controls how much small clusters are penalized. 
In general, we think that a slight over-estimation of the true number of clusters is not an issue of major concern in the context of patient stratification, since in this case, the primary interest typically lies in the characterization and the interpretation of clusters rather than the inference on the exact number of underlying clusters. As has been noted by \citet{onogi2011DPncluster}, even though the extra clusters produced by DPMM are considered as redundant and interpreted as over-estimation in simulation studies, they may provide useful information in real data applications since they reflect some relatively subtle heterogeneity that might be clinically interesting. In practice, in order to achieve better interpretability of the clustering results, the model-based ``representative'' clustering should be coupled with practitioners' subject-matter knowledge when determining the optimal number of clusters. If the inspection of the output indicates that the small clusters produced by DPMM are not of much clinical interest because they do not reflect a general pattern, the other larger and more representative clusters will be given more emphasis in the interpretation. In addition, if the profiles of some small clusters are very similar to those of some much larger clusters in terms of clinical meaningfulness, we can merge them \textit{a posteriori}. \textcolor{black}{From our point of view, this would be preferable to a method of low granularity, in which case some useful information on the population heterogeneity may be overlooked.}

The work in this paper has raised new research questions that are worthy of further investigation. Although our method is primarily developed for %stratifying patients into subgroups (in multi-arm trials) in 
the setting with a univariate and continuous target outcome, it can be extended to the multi-outcome setting. %with multiple and potentially correlated continuous outcomes. %The current implementation of BART in R (the \texttt{BART} package) cannot handle multivariate outcomes directly (and correlation might be an issue if we fit BART models separately for each outcome), but we can first ``decorrelate'' multiple outcomes by deriving uncorrelated principal components (PC) from the observed outcome data and run BART separately on each PC. Then we can re-transform the output from BART (on the ``PC scale'') to recover the potential outcomes on the ``original outcome scale'', which can subsequently be fed into the second-stage profile regression model. Preliminary simulation results (not presented in this paper) suggest that such a method extension performs well (provides sensible clustering/patient stratification results) in the setting with multiple continuous outcomes.

%The uncertainty associated with the clustering results produced by profile regression can be directly quantified under the Bayesian framework (Section \ref{sec:post_process}). However, 
Since we use the posterior mean of the potential outcome predictions (from BART) rather than the full posterior samples as the outcome in the profile regression model, the uncertainties associated with the predicted potential outcomes (the first stage) are not carried forward to the second stage and thus not reflected in the final output. We may employ the idea of Markov melding (a generic Bayesian computational method for evidence synthesis) to allow uncertainty propagation \citep{goudie2019markovmelding}.

Numerical experiments suggest that our proposed patient stratification method does not scale well to large datasets, %\textcolor{black}{(e.g. all male/female donors in the INTERVAL trial)}, 
mainly due to the computational inefficiency \textcolor{black}{of the current implementation of profile regression with a multivariate normal outcome in the \texttt{PReMiuMar} package.} %the second stage where profile regression is implemented for a multivariate continuous outcome. 
MCMC sampling can be prohibitively slow, and a variational inference algorithm for Dirichlet process mixture models may be considered to speed up the computation \citep{blei2006variational}.

In the profile regression model, both covariates ($\BX$) and outcome ($\BYstar$) inform the clustering structure. If the dimension of $\BX$ is much higher than that of $\BYstar$, the contribution of $\BYstar$ to the likelihood is likely to be overwhelmed by that of $\BX$ (i.e. covariates might dominate the likelihood and the relative contribution of the outcome may be undermined). Consequently, the impact of the response data on the cluster allocation will be small and the resulting clusters will be formed mainly based on the similarity in the covariate space. \citet{bigelow2009jointcluster} argued that this might be a desirable property for some epidemiological studies. However, in some other cases where we expect the outcome to play a more important role in the clustering, we might need to upweight the outcome likelihood. In practice, the weight may be subjective and may depend heavily on the contexts and research aims.

%  The \backmatter command formats the subsequent headings so that they
%  are in the journal style.  Please keep this command in your document
%  in this position, right after the final section of the main part of 
%  the paper and right before the Acknowledgements, Supporting Information (Supplementary %  Materials),   and References sections. 

\backmatter

%  This section is optional.  Here is where you will want to cite
%  grants, people who helped with the paper, etc.  But keep it short!

\section*{Acknowledgments}

This work was supported by the UK Medical Research Council programme MC\_UU\_00002/2 and the Cambridge International Scholarship. Participants in the INTERVAL trial were recruited with the active collaboration of NHS Blood and Transplant England (\url{www.nhsbt.nhs.uk}), which has supported field work and other elements of the trial. 
The academic coordinating centre for INTERVAL was supported by core funding from: NIHR Blood and Transplant Research Unit in Donor Health and Genomics (NIHR BTRU-2014-10024), UK Medical Research Council (MR/L003120/1), British Heart Foundation (SP/09/002; RG/13/13/30194; RG/18/13/33946) and the NIHR [Cambridge Biomedical Research Centre at the Cambridge University Hospitals NHS Foundation Trust]. A complete list of the investigators and contributors to the INTERVAL trial is provided in \citet{angelantonio2017interval}. The academic coordinating centre would like to thank blood donor centre staffs and blood donors for participating in the INTERVAL trial. 
This work was also supported by Health Data Research UK, which is funded by the UK Medical Research Council, Engineering and Physical Sciences Research Council, Economic and Social Research Council, Department of Health and Social Care (England), Chief Scientist Office of the Scottish Government Health and Social Care Directorates, Health and Social Care Research and Development Division (Welsh Government), Public Health Agency (Northern Ireland), British Heart Foundation and Wellcome. The views expressed in this paper are those of the authors and not necessarily those of the NHS, the NIHR or the Department of Health and Social Care.

%  Here, we create the bibliographic entries manually, following the
%  journal style.  If you use this method or use natbib, PLEASE PAY
%  CAREFUL ATTENTION TO THE BIBLIOGRAPHIC STYLE IN A RECENT ISSUE OF
%  THE JOURNAL AND FOLLOW IT!  Failure to follow stylistic conventions
%  just lengthens the time spend copyediting your paper and hence its
%  position in the publication queue should it be accepted.

%  We greatly prefer that you incorporate the references for your
%  article into the body of the article as we have done here 
%  (you can use natbib or not as you choose) than use BiBTeX,
%  so that your article is self-contained in one file.
%  If you do use BiBTeX, please use the .bst file that comes with 
%  the distribution.  In this case, replace the thebibliography
%  environment below by 
%

\bibliographystyle{biom}
\bibliography{stratification}

\section*{Supporting Information}
Additional supporting information may be found online in the Supporting Information section at the end of the article. The R code for implementing the proposed patient stratification method is available at \url{https://github.com/yx299/stratification}.
%\vspace*{-8pt}

\label{lastpage}

\end{document}